\documentclass[aps,prl,twocolumn,showpacs]{revtex4}
\usepackage{amsfonts,amsmath,amssymb,graphicx}
\usepackage{amsfonts}
\usepackage{amsmath}
\usepackage{amssymb}
\usepackage{graphicx}

\def\kbar{\protect\@kbar}
\def\@kbar{\relax \bgroup
\def\@tempa{\hbox{\raise.73\ht0
\hbox to0pt{\kern.25\wd0\vrule width.5\wd0 height.1pt
depth.1pt\hss}\box0}}\mathchoice{\setbox0\hbox{$\displaystyle
k$}\@tempa}{\setbox0\hbox{$\textstyle
k$}\@tempa}{\setbox0\hbox{$\scriptstyle
k$}\@tempa}{\setbox0\hbox{$\scriptscriptstyle k$}\@tempa}\egroup}

\begin{document}

\title{\textbf{Topologically Universal Spectral Hierarchies of Quasiperiodic
Systems }}
\author{Itzhack Dana}
\affiliation{Department of Physics, Bar-Ilan University, Ramat-Gan 52900, Israel}

\begin{abstract}
Topological properties of energy spectra of general one-dimensional
quasiperiodic systems, describing also Bloch electrons in magnetic fields, are
studied for an infinity of irrational modulation frequencies corresponding
to irrational numbers of flux quanta per unit cell. These frequencies
include well-known ones considered in works on Fibonacci quasicrystals. It
is shown that the spectrum for any such frequency exhibits a self-similar
hierarchy of clusters characterized by universal (system-independent) values
of Chern topological integers which are exactly determined. The cluster
hierarchy provides a simple and systematic organization of all the spectral
gaps, labeled by universal topological numbers which are exactly determinable, thus avoiding their numerical evaluation using rational approximants of the irrational frequency. These numbers give both the quantum Hall conductance of the system and the winding number of the edge-state energy traversing a gap as a Bloch quasimomentum is varied.
\end{abstract}

\maketitle
  
\begin{center}
\textbf{I. INTRODUCTION}
\end{center}

The topological characterization of band spectra was introduced in the
seminal paper by Thouless, Kohmoto, Nightingale, and den Nijs (TKNN) \cite{tknn} to explain the quantum Hall effect in a two-dimensional (2D) periodic
potential. This characterization was subsequently studied in detail and
extended to many other systems \cite{ass,bs,ahm,mk,daz,dz,jz0,id0,hk,yh,qc1,qc2,qc3,qc4,qc5,qc6,qhg,kz,kz1,kz2}.
TKNN considered particular models of Bloch electrons in \textquotedblleft
rational\textquotedblright\ magnetic fields with flux $\phi =\phi _{0}p/q$
per unit cell, where $\phi _{0}=hc/e$ is the quantum of flux and $(p,q)$ are
coprime integers. They showed that a magnetic band $b$ is characterized by
an integer, here denoted by $\sigma _{b}$, giving the contribution $\sigma
_{b}e^{2}/h$ of the band to the quantized Hall conductance of the system in
linear-response theory. This integer is a Chern topological invariant for
the band \cite{ass,bs} and satisfies the Diophantine equation \cite{tknn,ahm,daz,dz}: 
\begin{equation}
p\sigma _{b}+q\mu _{b}=1,  \label{de}
\end{equation}
where $\mu _{b}$ is a second integer. It was later shown \cite{daz,dz} that
Eq. (\ref{de}) is a general result which follows just from magnetic
(phase-space) translational invariance \cite{jz,dz0} and reflects the 
$q$-fold degeneracy of a magnetic band \cite{dz}. Summing Eq. (\ref{de}) over $N
$ bands, with the Fermi energy lying in the gap between the $N$th and $(N+1)$th bands, leads to the second general result of work \cite{daz}:
\begin{equation}
\varphi \sigma +\mu =\rho ,  \label{thc}
\end{equation}
where $\varphi =\phi /\phi _{0}=p/q$, $\rho =N/q$ is the number of electrons per unit cell, and $(\sigma ,\mu )$ are topological integers having the following meaning: $\sigma e^{2}/h$ is the quantum Hall conductance of the system \cite{daz,kunz} and $\mu e$ is the charge per unit cell that is transported when the periodic potential is dragged adiabatically by one lattice constant \cite{kunz}. A significant advantage of Eq. (\ref{thc}) over Eq. (\ref{de}) is that it can be immediately extended to irrational $\varphi $ \cite{daz}, by taking the limit of $p$, $q\rightarrow \infty $. Eq. (\ref{de}) is not defined
in this limit since a band reduces to an infinitely degenerate level \cite{rwo}. For irrational $\varphi$ and for $\rho $ corresponding to a gap, Eq. (\ref{thc}) has only one solution $(\sigma ,\mu )$, which is thus \emph{universal} (system independent). In contrast, for rational $\varphi $ and $\rho$ in a gap, Eq. (\ref{thc}) has an infinite number of solutions $(\sigma '+lq,\mu '-lp)$, where $(\sigma ',\mu ')$ is some solution and $l$ is any integer. In fact, the value of $\sigma$ (or $\mu$) for rational $\varphi $ is system dependent \cite{tknn,ahm,id0,hk} and changes generically by $\pm q$ (or $\mp p$) at band degeneracies \cite{bs,hk,qc2,qc5}. 

It is then natural to ask whether and how one can determine the universal topological numbers of gaps, systematically organized in some spectral hierarchy for irrational $\varphi $, without using rational approximants of $\varphi $.   This question is most relevant also in the broader context of general one-dimensional (1D) quasiperiodic systems. In fact, it is now well established that effective Hamiltonians for Bloch electrons in magnetic fields \cite{dz,hm,ar,dh} can describe, for irrational $\varphi $, a large class of 1D quasiperiodic systems \cite{kz1,kz2,hm,ar,dh}, ranging from Harper models or generalized Harper models \cite{hm,ar,dh,aa,djt,cey,hthk,aj,tb0} to quite different systems such as Fibonacci quasicrystals \cite{tb0,tb1,tb2,tb3,tb4,tb5,tb6,tb7,tb8,tb9,tb10}. Then, remarkable spectral structures of Fibonacci quasicrystals \cite{tb4,tb5,tb6,tb7,tb8,tb9} can be exhibited by Bloch electrons in a magnetic field and the gaps in these structures can thus be labeled by topological numbers $(\sigma ,\mu )$. Besides giving the quantum Hall conductance, $\sigma $ is the winding number of the edge-state energy \cite{yh} traversing the gap $(\sigma ,\mu )$ of the 1D system as a parameter is varied. Recently \cite{kz}, this phenomenon has been experimentally observed.

In this paper, we study topological properties of the energy
spectra of general 1D quasiperiodic systems, describing also Bloch electrons in
magnetic fields, for an infinity of irrational values of $\varphi $. These
values correspond to quasiperiodicity frequencies including
well-known frequencies assumed in studies of Fibonacci quasicrystals. We
show that for any such value of $\varphi $ the energy spectrum exhibits a
self-similar hierarchy of clusters characterized by universal values of
Chern integers which are exactly determined. This cluster hierarchy provides
a simple and systematic organization of all the spectral gaps, labeled by universal topological numbers which are exactly determinable, thus avoiding their numerical evaluation using rational approximants of $\varphi $. Smaller gaps with generally larger values of topological numbers are found in higher generations of the hierarchy. 

Sec. II is a brief summary of known facts about effective Hamiltonians for Bloch electrons in magnetic fields. The main results appear in Sec. III. Examples are given in Sec. IV and conclusions are presented in Sec. V.   

\begin{center}
\textbf{II. EFFECTIVE HAMILTONIANS AND 1D QUASIPERIODIC SYSTEMS}
\end{center}

It is well known \cite{dz,ar,dh} that for rational $\varphi =p/q$ and for a 2D periodic potential $V(x,y)$ that is weak relative to the Landau-level spacing, a Landau level splits into $p$ magnetic bands.  The energy spectrum $E$ of these bands can be shown to be the spectrum of an effective Hamiltonian $\hat{H}_{\mathrm{eff}}$, an operator which is derived from $V(x,y)$ and whose eigenvalue problem can be expressed as a difference equation in some representation. For example, in the simple case of $V(x,y)=v_{1}V_1(x)+v_{2}\cos (y)$, where $V_1(x)$ is an arbitrary $2\pi$-periodic function and $v_{1}$ and $v_{2}$ are suitably chosen coefficients, the difference equation is
\begin{equation}
\psi _{n+1}+\psi _{n-1}+\lambda V_1(2\pi n\nu +k)\psi _{n}=E\psi _{n}.
\label{tbc}
\end{equation}
Here $\psi _{n}$ is a representation of the magnetic Bloch states, $\lambda$ is an arbitrary parameter, and $k$ is a Bloch quasimomentum. Equation (\ref{tbc}) describes a tight-binding chain with modulation frequency $\nu =1/\varphi$. For irrational $\nu$, this chain is a 1D quasiperiodic system. Extreme cases of this system are the Harper model with $V_1(x)=\cos (x)$ and the Fibonacci quasicrystal with $V_1(x)=\left[
\left\lfloor x/(2\pi )+2\nu \right\rfloor -\left\lfloor x/(2\pi )+\nu
\right\rfloor \right] -1$, where $\left\lfloor \cdot \right\rfloor $ is the
floor function. Much more general periodic potentials $V(x,y)$ lead to a large class of 1D quasiperiodic systems \cite{dz,kz1} in which $\nu $ still appears only in the argument $2\pi n\nu $ of $2\pi $-periodic functions as in Eq. (\ref{tbc}). Then, one can replace $\nu $ by $\left\lfloor \nu \right\rfloor $, i.e., one can assume that $\nu <1$.

In the regime of strong periodic potential relative to the Landau-level
spacing and for $\varphi =p/q$, a Bloch band ``splits" into $q$ magnetic bands whose energy spectrum is that of an effective Hamiltonian $\hat{H}_{\mathrm{eff}}$, derived from the Bloch-band function using the Peierls substitution \cite{hm,dh}. For irrational $\varphi$, the eigenvalue problem for $\hat{H}_{\mathrm{eff}}$ is again described by a 1D quasiperiodic system but with modulation frequency $\nu =\varphi$. For the sake of definiteness and without loss of generality, we shall assume in what follows the regime above of weak periodic potential.

\begin{center}
\textbf{III. TOPOLOGICALLY UNIVERSAL SPECTRAL HIERARCHIES}
\end{center}

Consider the $p$ magnetic bands splitting from one Landau level for $\varphi =p/q$. Summing Eq. (\ref{de}) over a cluster of $N$ neighboring bands, $N\leq p$, we see that the cluster is characterized by topological integers $(\sigma ,\mu )$ satisfying the Diophantine equation:
\begin{equation}
\sigma +\nu \mu =\eta ,  \label{cde}
\end{equation}
where $\sigma e^{2}/h$ is the contribution of the cluster to the total Hall
conductance $e^{2}/h$ of the Landau level, $\nu =1/\varphi$ is the modulation frequency (see Sec. II), and $\eta =N/p$ is the spectral occupation fraction (SOF) of the cluster. Equation (\ref{cde}) extends straightforwardly to irrational $\varphi $ or $\nu$, as in the case of Eq. (\ref{thc}). As mentioned in Sec. II, we can restrict our attention to frequencies $\nu <1$, without loss of generality. We shall consider irrational values of $\nu <1$ that are the positive root
of the equation
\begin{equation}
m\nu +\nu ^{2}=1,  \label{enu}
\end{equation}
for arbitrary positive integer $m$. Explicitly, $\nu $ and its
continued-fraction expansion are given by
\begin{equation}
\nu =\frac{\sqrt{m^{2}+4}-m}{2}=[0,m,m,m,...].  \label{num}
\end{equation}
Well-known frequencies (\ref{num}) considered in works on Fibonacci
quasicrystals \cite{tb4,tb5,tb6,tb7,tb8,tb9}\ are the inverse of the golden
mean ($m=1$), of the silver mean ($m=2$), and of the bronze mean ($m=3$).
The $s$th rational approximant $\nu _{s}=q_{s}/p_{s}$ of $\nu $, $s\geq 1$,
is obtained by truncating the continued-fraction expansion (\ref{num}) at
the $s$th stage. One then gets: $q_{s}=F_{s-1}$ and $p_{s}=F_{s}$, where 
$F_{s}$ are the generalized Fibonacci numbers \cite{tb7} satisfying the
recursion relation
\begin{equation}
F_{s}=mF_{s-1}+F_{s-2},\ \ \ \ s>0,  \label{gf}
\end{equation}
with initial conditions $F_{-1}=1$ and $F_{0}=0$.

In order to get a full topological characterization of the spectrum, we
assume from now on that all the spectral gaps for the frequencies $\nu
_{s}=F_{s-1}/F_{s}$ (arbitrary $s$, including $\nu _{\infty }=\nu $) are
open. This is known to hold at least for generalized Harper models (\ref{tbc}) with smooth $V$ \cite{cey,aj} and for the Fibonacci quasicrystal \cite{tb10} in some parameter range. Relation (\ref{gf}) then clearly shows that
the $F_{s}$ isolated bands for $\nu =\nu _{s}$ can be naturally grouped into 
$m+1$ clusters: $m$ clusters, each with $F_{s-1}$ bands and SOF $\eta
_{1}=F_{s-1}/F_{s}$, and one cluster with $F_{s-2}$ bands and SOF $\eta
_{2}=F_{s-2}/F_{s}$. To remove some arbitrariness in this grouping of the 
$F_{s}$ bands, we impose a convenient \emph{energy ordering} of the $m+1$
clusters: The cluster with SOF $\eta _{2}$ consists of the $F_{s-2}$ bands
that are the highest in energy, i.e., the energy of this cluster is above
that of all the $m$ clusters with SOF $\eta _{1}$. The $m+1$ clusters with
this energy ordering define the first generation of a hierarchy. In the
second generation, each of these clusters splits into $m+1$ subclusters
according to $F_{s-1}=mF_{s-2}+F_{s-3}$ (for SOF $\eta _{1}$) or 
$F_{s-2}=mF_{s-3}+F_{s-4}$ (for SOF $\eta _{2}$) with energy ordering similar
to the above one. This process can be continued up to generation 
$\bar{g}=\left\lfloor (s-1)/2\right\rfloor $.

Taking now the limit of $s,\bar{g}\rightarrow \infty $, we see that $\eta
_{1}\rightarrow \nu $, $\eta _{2}\rightarrow \nu ^{2}$, $F_{s-3}/F_{s}
\rightarrow \nu ^{3}$, $F_{s-4}/F_{s}\rightarrow \nu ^{4}$, etc. We then get
for irrational $\nu $ an infinite hierarchy of generations of clusters as
follows: In the first ($g=1$) generation, one has $m$ clusters $\mathcal{C}
_{j}$ with SOF $\eta _{j}=\nu $ each, $j=1,\dots ,m$, and, above them in
energy, one cluster $\mathcal{C}_{m+1\text{ }}$with SOF $\eta _{m+1}=\nu
^{2} $; the \textquotedblleft resolution of the identity\textquotedblright\ 
$\sum_{j=1}^{m+1}\eta _{j}=1$ is guaranteed by Eq. (\ref{enu}). For any fixed 
$j_{1}=1,\dots ,m+1$, a cluster $\mathcal{C}_{j=j_{1}}$ splits into $m+1$
subclusters $\mathcal{C}_{j_{1},j_{2}}$ in generation $g=2$, with SOFs $\eta _{j_{1},j_{2}}=\eta _{j_{1}}\eta _{j_{2}}$, $j_{2}=1,\dots ,m+1$; again,
the energy of $\mathcal{C}_{j_{1},m+1}$ is above that of $\mathcal{C}
_{j_{1},j_{2}}$, $j_{2}=1,\dots ,m$. In general, the $g$th generation
consists of the $(m+1)^{g}$ \textquotedblleft elementary\textquotedblright\
clusters $\mathcal{C}_{j_{1},\dots ,j_{g}}$ with SOFs
\begin{equation}
\eta _{j_{1},\dots ,j_{g}}=\prod\limits_{l=1}^{g}\eta _{j_{l}}=\nu ^{c},\ \
\ g\leq c\leq 2g,  \label{ss}
\end{equation}
for $j_{l}=1,\dots ,m+1$ and $l=1,\dots ,g$. The resolution of the identity 
$\sum_{j_{1},\dots ,j_{g}=1}^{m+1}\eta _{j_{1},\dots ,j_{g}}=1$ is just the 
$g$th power of Eq. (\ref{enu}). This hierarchy is self-similar in the sense
that each elementary cluster in generation $g$ always splits into $m+1$
subclusters in generation $g+1$ with an energy ordering similar to that in generation $g$. Also, according to Eq. (\ref{ss}), the SOFs of the $m+1$ subclusters are obtained by scaling the SOF of the parent cluster with the simple factor $\eta _{j_{g+1}}=\nu $ or $\nu ^{2}$.

Let us show that the elementary clusters with SOF $\eta =\nu ^{c}$ have well
defined Chern integers $(\sigma _{c},\mu _{c})$. We first derive a formula
for $\nu ^{c}$ in terms of the generalized Fibonacci numbers $F_{s}$. Using the fact that $\nu $ and $-1/\nu $ are the two roots of Eq. (\ref{enu}), it is easy to check that $F_{s}=a\nu ^{-s}+d(-\nu )^{s}$ satisfies Eq. (\ref{gf}) for some
constants $a$ and $d$ that are determined from $F_{-1}=1$ and $F_{0}=0$. We
get: 
\begin{equation}
F_{s}=\frac{\nu ^{-s}-(-\nu )^{s}}{\sqrt{m^{2}+4}}.  \label{fs}
\end{equation}
Writing Eq. (\ref{fs}) for $s=c$ and $s=c-1$, we can then extract the formula
for $\nu ^{c}$: 
\begin{equation}
\nu ^{c}=(-1)^{c}(F_{c-1}-\nu F_{c}).  \label{nuc}
\end{equation}
Using formula (\ref{nuc}) and Eq. (\ref{cde}) with $\eta =\nu ^{c}$, $\sigma
=\sigma _{c}$, and $\mu =\mu _{c}$, we obtain:
\begin{equation}
\sigma _{c}=(-1)^{c}F_{c-1},\ \ \ \ \mu _{c}=\sigma _{c+1}=(-1)^{c+1}F_{c},
\label{smc}
\end{equation}
where $\sigma _{0}=F_{-1}=1$ corresponds to the entire Landau level with $\eta =1$. Equations (\ref{smc}) give the universal, system-independent
values of the Chern integers of the elementary clusters for frequency (\ref{num}). Remarkably, Eqs. (\ref{smc}) do not depend explicitly on $m$, only
implicitly through $F_{c-1}$ and $F_{c}$. Using Eqs. (\ref{gf})
and (\ref{smc}), we get the following recursion relations between the Chern integers $(\sigma _{c},\mu _{c})$ of elementary clusters with SOFs $\eta =\nu ^{c}$: 
\begin{equation}
\sigma_{c+1}=\sigma_{c-1}-m\sigma_c,\ \ \
\mu_{c+1}=\mu_{c-1}-m\mu_c.  \label{smj}
\end{equation}
For large $m$, Eqs. (\ref{smj}) connect, in most cases, topological numbers in one generation with those in the two previous generations.

All the general results above are illustrated in Fig. 1 for the silver-mean case of $\nu =\sqrt{2}-1$ ($m=2$). 
\begin{figure}[tbp]
\includegraphics[width=8.0cm]{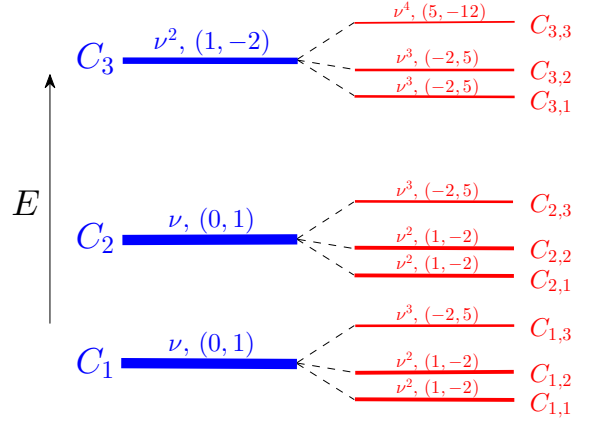}
\caption{(Color online) Schematic illustration of the spectral hierarchy in the silver-mean case of $\nu =\sqrt{2}-1$ ($m=2$), showing the first two generations. In the first generation, there are $m+1=3$ elementary clusters $\mathcal{C}_j$, $j=1,2,3$ (thick blue segments). As indicated, both $\mathcal{C}_1$ and $\mathcal{C}_2$ have SOF $\eta =\nu$ and Chern integers $(\sigma ,\mu )=(0,1)$, while $\mathcal{C}_3$ has $\eta =\nu^2$, $(\sigma ,\mu )=(1,-2)$, and energy higher than that of $\mathcal{C}_1$ and $\mathcal{C}_2$. In the second generation, each cluster $\mathcal{C}_j$ splits into three subclusters $\mathcal{C}_{j,j'}$, $j'=1,2,3$ (thinner red segments), whose SOFs $\eta$ and Chern integers $(\sigma ,\mu )$ are indicated. Again, the third subcluster has characteristics different from the first two ones and is higher in energy than them. The thickness of each segment qualitatively represents the value of the SOF $\eta$ for the corresponding cluster.}
\label{fig1}
\end{figure}

An arbitrary, generally non-elementary cluster with given SOF $\eta \leq 1$
is composed of elementary clusters with SOFs $\nu ^{c}$, $c\geq 1$.
One can express $\eta $ in terms of $\nu ^{c}$ by expanding $\eta $ in the
non-integer basis \cite{nib} $\nu ^{-1}$:
\begin{equation}
\eta =\sum_{c=1}^{\infty }r_{c}\nu ^{c},  \label{ee}
\end{equation}
where the \textquotedblleft digits\textquotedblright\ $r_{c}$ are integers
which range in the interval $0\leq r_{c}\leq \left\lfloor \nu
^{-1}\right\rfloor =m$; $r_{c}$ is the number of elementary clusters with
SOF $\nu ^{c}$ in the given non-elementary cluster. The integers 
$r_{c} $ are determined by the following algorithm \cite{nib}:
\begin{equation}
r_{c}=\left\lfloor \chi _{c}/\nu \right\rfloor ,\ \ \ \ \chi _{c}=\chi
_{c-1}/\nu -\left\lfloor \chi _{c-1}/\nu \right\rfloor  \label{ga}
\end{equation}
for $c>1$ and $r_{1}=\left\lfloor \eta /\nu \right\rfloor $, $\chi _{1}=\eta 
$. Since the SOF $\nu ^{c}$ is associated with Chern integers (\ref{smc}),
the cluster with SOF (\ref{ee}) has the formal topological characterization:
\begin{equation}
\sigma =\sum_{c=1}^{\infty }(-1)^{c}r_{c}F_{c-1},\ \ \ \mu
=\sum_{c=1}^{\infty }(-1)^{c+1}r_{c}F_{c}.  \label{sm}
\end{equation}
Thus, if the sum in Eq. (\ref{ee}) contains a finite number of terms, as it will be required below, $(\sigma ,\mu )$ in Eqs. (\ref{sm}) exist and the cluster is topologically well defined. For rational $\eta $, as well as for an infinity of irrational values of $\eta $, $(\sigma ,\mu )$ do not exist; see note \cite{note}.

A gap in some generation of the hierarchy is labeled by universal
topological numbers $\sigma $ and $\mu $ given by the sum of $\sigma _{c}$
and $\mu _{c}$, respectively, for all the energy-ordered elementary clusters in
that generation below the gap.

We now show how to determine the precise location in the hierarchy of any
gap in the spectrum. The gap is defined by a filling factor, i.e., the SOF 
$\eta $ of a generally non-elementary cluster starting from the bottom of the
Landau level and above which the gap lies. As we shall see, the gap will be
located in a well-defined (finite) generation $g$ only if the expansion 
(\ref{ee}) for $\eta $ is finite; we denote by $\bar{c}$ the largest value of $c$ in (\ref{ee}). Then, the location of the gap in the hierarchy, with the
given energy ordering of the elementary clusters, is determined from this
finite expansion as follows. Let us form the sequence $j_{1},j_{2},\dots ,j_{
\bar{c}}$, where $j_{c}=r_{c}+1$ for $c<\bar{c}$ and $j_{\bar{c}}=r_{\bar{c}}$. Every time that $r_{c}=m$ for $c<\bar{c}$ one must necessarily have
$r_{c+1}=0$ from the algorithm (\ref{ga}). We delete from the sequence above
all elements with $r_{c}=0$ ($j_{c}=1$) if $r_{c-1}=m$ ($j_{c-1}=m+1$), thus
obtaining the (usually shorter) sequence $j_{1},j_{2},\dots ,j_{g}$, $g\leq 
\bar{c}$. It is then easy to see that the gap is located just above the
elementary cluster $\mathcal{C}_{j_{1},j_{2},\dots ,j_{g}}$ in the $g$th
generation of the hierarchy. The topological numbers $(\sigma ,\mu )$
labeling the gap are obtained from the expansions (\ref{sm}) using the
integers $r_{c}$ determined from the given value of $\eta $ by the algorithm
(\ref{ga}).

Due to Eqs. (\ref{ss}), (\ref{smc}), and (\ref{smj}), the absolute values of
the Chern integers of the clusters and of the spectral gaps have a generally
increasing trend in successive generations of the hierarchy.

\begin{center}
\textbf{IV. EXAMPLES}
\end{center}

We show here how the topologically universal spectral hierarchy in the golden-mean case of $\nu =(\sqrt{5}-1)/2$ ($m=1$) is exhibited by two systems having significantly different spectra. These are the Harper model and the Fibonacci quasicrystal given by the quasiperiodic chain (\ref{tbc}) with two quite different functions $V_1(x)$ (see Sec. II). Figs. 2 and 3 show the first four generations, or parts of them, of the topologically universal hierarchy in the spectra of the two systems for $\lambda =2$. The relevant values of cluster SOFs and topological numbers of spectral gaps were exactly determined from the general results in Sec. III. The plotted spectra are the band spectra for the rational approximant $34/55$ of $\nu $. The elementary clusters in each of the four generations were identified as the corresponding band clusters for this approximant; see the definition of such clusters at the beginning of Sec. III. We have checked that all the gaps between the band clusters are indeed open and are stable, i.e., they essentially do not change for higher-order approximants. In Figs. 2(a) and 3(a) all the spectrum is shown and the clusters $\mathcal{C}_{1}$ and $\mathcal{C}_{2}$ in the first generation are indicated by boxes. The other figures show three successive zooms of the cluster $\mathcal{C}_{1}$.
\begin{figure}[tbp]
\includegraphics[width=8.0cm]{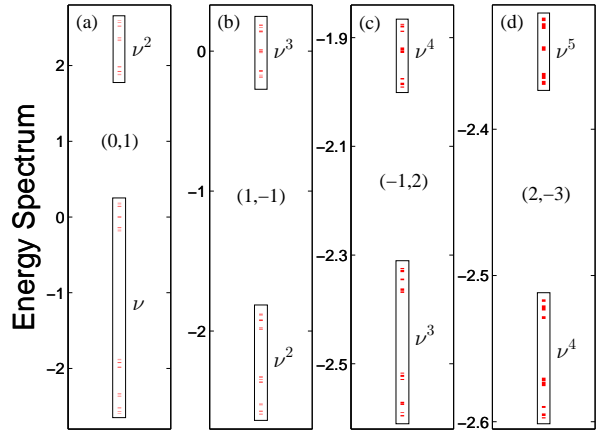}
\caption{(Color online) (a) Energy spectrum (red thick segments) of the
Harper model [Eq. (\protect\ref{tbc}) with $V_1(x)=\cos (x)$] for $\protect\nu
=(\protect\sqrt{5}-1)/2$ and $\protect\lambda =2$, plotted using the
rational approximant $34/55$ of $\protect\nu $. The two boxes define the
$m+1=2$ clusters in the first generation of the spectral hierarchy, with
indication of their SOFs, $\protect\nu $ and $\protect\nu ^{2}$, and
topological numbers $(\protect\sigma ,\protect\mu )=(0,1)$ of the gap
between them. Plots (b), (c), and (d) show parts of the second, third, and
fourth generation of the spectral hierarchy, obtained by zooming the lower
cluster in (a), (b), and (c), respectively; the notation is as in (a). }
\label{fig2}
\end{figure}

\begin{figure}[tbp]
\includegraphics[width=8.0cm]{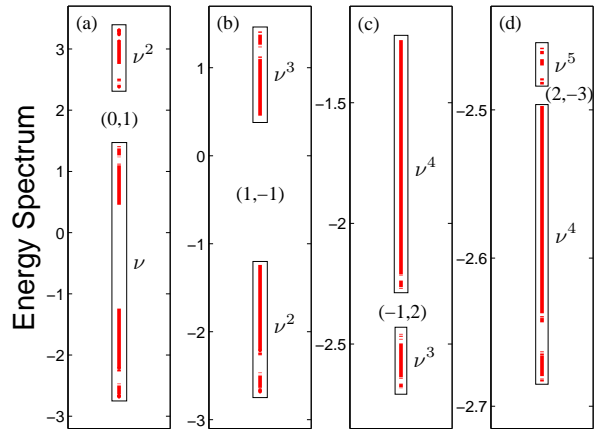}
\caption{(Color online) Similar to Fig. 2 but for the Fibonacci
quasicrystal, i.e., Eq. (\protect\ref{tbc}) with $V_1(x)=\left[ \left\lfloor
x/(2\protect\pi )+2\protect\nu \right\rfloor -\left\lfloor x/(2\protect\pi )+
\protect\nu \right\rfloor \right] -1$. In the calculations, a very accurate
smooth approximation of $V_1(x)$ was used, given by \protect\cite{kz1}
$V_1(x)\approx \tanh \left\{ \protect\beta \left[ \cos \left( x+3\protect\pi 
\protect\nu \right) -\cos (\protect\pi \protect\nu )\right] \right\} /\tanh 
(\protect\beta )$, with $\protect\beta =100$.}
\label{fig3}
\end{figure}

The spectral hierarchy illustrated in Figs. 2 and 3 should be compared with
the well-known trifurcation hierarchy naturally exhibited by the Harper
model \cite{dh} and the Fibonacci quasicrystal \cite{tb4,tb5,tb6,tb7,tb8}.
The latter hierarchy is based on the band-cluster splitting with
$F_{s-2}+F_{s-3}+F_{s-2}=F_{s}$ (or $\nu ^{2}+\nu^{3}+\nu ^{2}=1$ for $s\rightarrow \infty$) and is clearly visible in Figs. 2(a) and 3(a). Of course, this is fully equivalent to our bifurcation ($m+1=2$) hierarchy with
$F_{s-1}+F_{s-2}=F_{s}$ or $\nu +\nu ^{2}=1$. The general results in Sec. III can be easily expressed, for $m=1$, in terms of the trifurcation hierarchy. This hierarchy, however, may not be a natural one for a general 1D quasiperiodic system. Therefore, for the sake of simplicity and definiteness, we adopt the $m+1$ hierarchy for arbitrary $m$ to all systems.

\begin{center}
\textbf{V. CONCLUSIONS}
\end{center}

In conclusion, we have exactly determined, apparently for the first time, the universal topological numbers of all spectral clusters and gaps, systematically organized in well-defined self-similar hierarchies, for general 1D quasiperiodic systems with irrational modulation frequencies (\ref{num}). These frequencies include well-known ones considered in previous works. In general, it difficult to calculate numerically the universal topological numbers using the standard, system-dependent approach based on successive rational approximants of the irrational frequency. Our results straightforwardly provide the universal values of the quantum Hall conductances [or winding numbers of edge-state energies as a Bloch quasimomentum such as $k$ in Eq. (\ref{tbc}) is varied \cite{yh}] for a large class of interesting systems \cite{qhg,kz,kz1,kz2,hthk,aj,tb0,tb1,tb2,tb3,tb4,tb5,tb6,tb7,tb8,tb9,tb10}. It should be possible to extend our results to a set of irrational frequencies even larger than (\ref{num}).

If two systems have the same irrational frequency and can be continuously
deformed into each other, such as the Harper model and the Fibonacci quasicrystal \cite{kz1}, the topological numbers of a gap will not change
if this gap closes and re-opens during the deformation. However, such
changes (quantum phase transitions) will generally occur if the frequency is
varied. This was experimentally observed quite recently \cite{kz2} by
deforming a system with golden-mean frequency ($m=1$) to one with irrational
frequency not in the set (\ref{num}). It would be interesting to study, both
theoretically and experimentally, the nature of the quantum phase transition
when both the initial and final frequency belong to the set (\ref{num}).
This is a transition between two different universality classes of
topological numbers associated with the well-defined spectral hierarchies above.

\begin{center}
\textbf{ACKNOWLEDGMENTS}
\end{center}

The author has benefited from useful discussions with J.E. Avron, Y.E.
Kraus, G. Murthy, and O. Zilberberg.


\begin{thebibliography}{99}
\bibitem{note} The allowed values of $\eta $ for which $(\sigma ,\mu )$
exist must satisfy Eq. (\ref{cde}) with irrational $\nu $ and cannot be
therefore rational. As an explicit example of an infinite set of non-allowed
irrational values of $\eta $, consider $\eta =\sum_{c=0}^{\infty }\nu
^{j+2lc}$ for $j=0,1$ and all integers $l>1$. Using Eq. (\ref{nuc}), it is
easy to see that $\eta =\nu ^{j}/(\nu F_{2l}-F_{2l-1}+1)$. This value of
$\eta $ is always irrational and its expansion $\eta =\sum_{c=0}^{\infty }\nu
^{j+2lc}$ clearly implies divergent sums (\ref{sm}) for $\sigma $ and $\mu $.
\end{thebibliography}
\end{document}